\documentclass[sigconf]{acmart}

\AtBeginDocument{%
  \providecommand\BibTeX{{%
    \normalfont B\kern-0.5em{\scshape i\kern-0.25em b}\kern-0.8em\TeX}}}

\usepackage{booktabs,multirow}
\usepackage{subfigure} 
\usepackage{balance}
\usepackage{graphicx}
\usepackage{caption}
\usepackage{algorithm}
\usepackage{algpseudocode}  
\usepackage{orcidlink}

\copyrightyear{2024} 
\acmYear{2024} 
\setcopyright{rightsretained} 
\acmConference[KDD '24]{Proceedings of the 30th ACM SIGKDD Conference on Knowledge Discovery and Data Mining}{August 25--29, 2024}{Barcelona, Spain}
\acmBooktitle{Proceedings of the 30th ACM SIGKDD Conference on Knowledge
Discovery and Data Mining (KDD '24), August 25--29, 2024, Barcelona,
Spain}\acmDOI{10.1145/3637528.3671968}
\acmISBN{979-8-4007-0490-1/24/08}

\makeatletter
\gdef\@copyrightpermission{
  \begin{minipage}{0.3\columnwidth}
   \href{https://creativecommons.org/licenses/by-nc-sa/4.0/}{\includegraphics[width=0.90\textwidth]{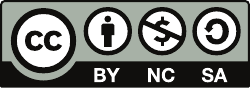}}
  \end{minipage}\hfill
  \begin{minipage}{0.7\columnwidth}
   \href{https://creativecommons.org/licenses/by-nc-sa/4.0/}{This work is licensed under a Creative Commons Attribution-NonCommercial-ShareAlike International 4.0 License.}
  \end{minipage}
  \vspace{5pt}
}
\makeatother

\begin{document}

\title{Predicting Long-term Dynamics of Complex Networks via Identifying Skeleton in Hyperbolic Space}

\author{Ruikun Li}
\affiliation{
  \institution{Shenzhen International Graduate School, Tsinghua University}
  \state{Shenzhen}
  \country{China}
}

\author{Huandong Wang}
\authornote{Huandong Wang is the corresponding author (wanghuandong@tsinghua.edu.cn).}
\affiliation{
  \institution{Department of Electronic Engineering\\ BNRist, Tsinghua University}
  \state{Beijing}
  \country{China}
}

\author{Jinghua Piao}
\affiliation{
  \institution{Department of Electronic Engineering\\ BNRist, Tsinghua University}
  \state{Beijing}
  \country{China}
}

\author{Qingmin Liao}
\affiliation{
  \institution{Shenzhen International Graduate School, Tsinghua University}
  \state{Shenzhen}
  \country{China}
}

\author{Yong Li}
\affiliation{
  \institution{Department of Electronic Engineering\\ BNRist, Tsinghua University}
  \state{Beijing}
  \country{China}
}

\renewcommand{\shortauthors}{Ruikun Li, Huandong Wang, Jinghua Piao, Qingmin Liao, and Yong Li}

\begin{abstract}
  Learning complex network dynamics is fundamental for understanding, modeling, and controlling real-world complex systems. Though great efforts have been made to predict the future states of nodes on networks, the capability of capturing long-term dynamics remains largely limited. This is because they overlook the fact that long-term dynamics in complex network are predominantly governed by their inherent low-dimensional manifolds, i.e., skeletons. Therefore, we propose the \underline{D}ynamics-\underline{I}nvariant \underline{Sk}eleton Neural \underline{Net}work (DiskNet), which identifies skeletons of complex networks based on the renormalization group structure in hyperbolic space to preserve both topological and dynamics properties. Specifically, we first condense complex networks with various dynamics into simple skeletons through physics-informed hyperbolic embeddings. Further, we design graph neural ordinary differential equations to capture the condensed dynamics on the skeletons. Finally, we recover the skeleton networks and dynamics to the original ones using a degree-based super-resolution module. Extensive experiments across three representative dynamics as well as five real-world and two synthetic networks demonstrate the superior performances of the proposed DiskNet, which outperforms the state-of-the-art baselines by an average of 10.18\% in terms of long-term prediction accuracy. Code for reproduction is available at: \href{https://github.com/tsinghua-fib-lab/DiskNet}{https://github.com/tsinghua-fib-lab/DiskNet}.
\end{abstract}

\ccsdesc[500]{Computing methodologies~Modeling methodologies}
\ccsdesc[300]{Computing methodologies~Learning latent representations}
\ccsdesc[300]{Applied computing~Physics}

\ccsdesc[500]{Computing methodologies~Machine learning}
\ccsdesc[500]{Networks~Network dynamics}
\ccsdesc[300]{Applied computing~Physics}

\keywords{Complex Network, Dynamical Systems, Graph Neural Networks, Neural ODE}


\maketitle

\section{Introduction}
The evolution behavior of numerous complex systems in the real world, such as the brain~\cite{chialvo2010emergent}, social networks~\cite{zanudo2017structure}, and ecosystems~\cite{hu2022emergent}, can be modeled as the dynamics on complex networks~\cite{barzel2013universality}
where the components within the system are regarded as nodes in the network, and the coupling interactions between components are regarded as edges~\cite{gao2016universal}. 
Learning the dynamics of these complex networks are instrumental to the ability to analyze them, facilitating numerous important applications including understanding their inherent resilience~\cite{gao2016universal}, predicting their future states~\cite{murphy2021deep,prasse2022predicting}, and controlling their states~\cite{liu2011controllability}, etc. However, a vast number of nodes and edges typically exist in complex networks. On the one hand, the large-scale network significantly increases the computational complexities of inference and estimation tasks on complex networks. On the other hand, the large number of less important nodes, acting as interference variables, potentially prevent us from uncovering the effective dynamics of the network~\cite{gao2016universal}. In this context, a fundamental question arises: given an arbitrary complex network, can we find a lower-dimensional skeleton of it, on which the coarse-scale long-term dynamics can be modeled with both high prediction performance and low computation cost?

As a long-standing problem, numerous methods have been developed to reduce the dimensionality of large-scale networks, including statistical physics methods~\cite{wang2023multi} and machine learning methods~\cite{jin2020graph, kumar2023featured}. In particular, the statistical physics methods utilize the renormalization group technique~\cite{garcia2018multiscale,villegas2023laplacian} to identify the skeleton that retains the degree distribution and clustering coefficient. The machine learning methods~\cite{jin2020graph, kumar2023featured} learn graph coarsening strategies that preserve the topological and node feature consistency between the original network and the network that is reconstructed using the coarsened network. All these methods only preserve the static topological and nodal information, which ignores the dynamics on the network~\cite{garcia2018multiscale,villegas2023laplacian,jin2020graph, kumar2023featured}. Consequently, they fail to capture the collective behavior of nodal dynamics to accurately forecast their long-term evolution.
Additionally, several analytical methods have shown that the dynamics of complex networks in terms of key properties, e.g.,  resilience~\cite{gao2016universal} or critical time delays~\cite{ma2023generalized}, can be embedded into a submanifold with ultra-low dimensionality~\cite{gao2016universal,ma2023generalized}.
However, they rely on overly simplified assumptions regarding network topology, e.g., homogeneous node degrees or degree-based mean-field~\cite{ma2023generalized,tu2021dimensionality}, and cannot obtain a skeleton with a desired dimensionality.
Taken together, whether the complex networks and their intricate dynamics can be condensed into low-dimensional skeletons is still an open problem.

Nevertheless, developing an effective approach to learn the skeleton and the corresponding dynamics on the skeleton network is also a difficult task with the following challenges.
First, identifying the skeleton that facilitates the most accurate predictions of the evolutionary behavior of complex networks is challenging.
The skeleton can be obtained by coarsening a cluster of nodes into a super-node. However, we lack sufficient understanding in terms of how to find such clusters while preserving the intrinsic dynamics in the obtained skeleton network.
Second, how to establish a reversible mapping between the node states in the skeleton network and the original network is the second challenge. 
It involves establishing an aggregating function to calculate the state of the super-node based on its sub-nodes, and a lifting function to map the state of each super-node back to its sub-nodes.
Since the aggregation process will introduce inevitably information loss, how to build an efficient lifting function is a crucial problem.

In this paper, we propose a novel deep learning framework, named \underline{D}ynamics-\underline{I}nvariant \underline{Sk}eleton Neural \underline{Net}work (DiskNet), to identify the low-dimensional skeleton of complex networks in hyperbolic space for modeling their long-term dynamics. This framework identifies the skeleton through physics-informed embeddings. These embeddings are initialized based on renormalization group techniques developed in hyperbolic space and then fine-tuned in the end-to-end training process, enabling the combination of knowledge and data to overcome the first challenge.
Then, DiskNet employs a graph neural ordinary differential equation to efficiently model the condensed dynamics on the skeleton network.
Finally, it utilizes an elaborately-designed degree-based super-resolution module as the lifting function to map the condensed dynamics back to the original network.
This module utilizes the homogeneity of nodes with respect to their degrees to identify clusters of nodes that share the same lifting network, thereby constructing an effective lifting function to address the second challenge.

Our contribution can be summarized as follows:
\begin{itemize}
    \item We propose to identify the skeleton based on hyperbolic embeddings inspired by physical knowledge from the renormalization group theory, enabling us to combine knowledge and data to learn an effective skeleton.
    \item We develop a powerful graph neural ordinary differential equations (ODEs) integrated with a novel super-resolution module utilizing the homogeneity of nodes with close degrees, allowing us to accurately model the condensed dynamics on the skeleton and map it to the original network.
    \item Extensive experimental results on three representative network dynamics on five real-world and two synthetic network topologies show that DiskNet outperforms state-of-the-art baselines by an average of 10.18\% in terms of prediction accuracy, indicating its superiority.
\end{itemize}
\section{Problem Formulation}
Without loss of generality, the network topology is represented by an adjacency matrix $A \in \{0, 1\}^{N \times N}$, where each element $a_{ij} = 1$ indicates the connection between nodes $i$ and $j$. Combining it with the node set $\mathcal{V}$, a complex network can be modeled as a graph $G = (\mathcal{V}, A)$. The network dynamics describe the evolution of node states on the graph $G$, taking into account the self-dynamics of each node and the interactions with its neighbors, following the equation 
\begin{equation}\label{equ:dynamics}
    \frac{dx_i}{dt} = f(x_i) + \sum^{N}_{j \neq i}{a_{ij}g(x_i,x_j)},
\end{equation}
where $f$ represents the self-dynamics and $g$ denotes the coupling dynamics between nodes. The node state matrix, denoted as $\textbf{X} \in \mathbb{R}^{N \times L \times d}$, represents the observed values of the $d$-dimensional states of $N$ nodes over a continuous sequence of $L$ time steps.

Furthermore, we consider the skeleton of the graph $G$ as $G_s = (\mathcal{V}_s, A_s)$, where $A_s \in \{0, 1\}^{\gamma N \times \gamma N}$ describes the connection between the super-nodes, which are defined as the aggregation of original nodes, in the set $\mathbb{V}_s$. Here, $\gamma$ represents the reduction ratio of the skeleton relative to the original topology, given by $\gamma = \frac{|\mathcal{V}_s|}{|\mathcal{V}|}$. The correspondence between the super-nodes and the original nodes is defined by the assignment matrix $P \in \{0, 1\}^{\gamma N \times N}$, where $p_{ij} = 1$ indicates that node $j$ is aggregated into super-node $i$. The function $x_{s,i}=u(x_{P_i})$ describes how the state of super-node $i$ is aggregated from its sub-nodes. The dynamics of the network on the skeleton, denoted as $\frac{dx_{s,i}}{dt}$, is defined in the same form as Eqn. \ref{equ:dynamics}. The state of a super-node is mapped back to its sub-nodes as $x_i = v(x_{s,i})$ by the lifting function $v$. We emphasize that obtaining the network skeleton through aggregation rather than pruning is aimed at consolidating the collective dynamics of similar nodes into a representative super-node.

\begin{figure}[!t]
\centering
\includegraphics[width=0.48\textwidth]{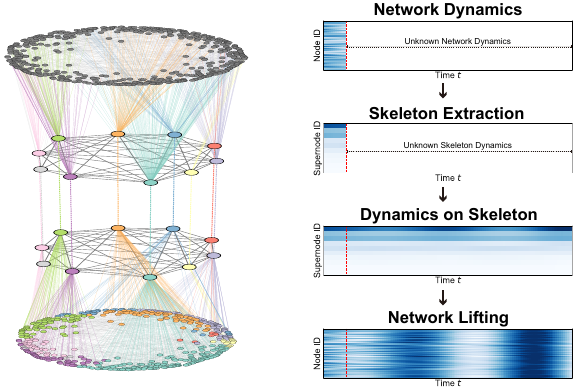}
\caption{Long-term prediction via the skeleton of complex network dynamics.}
\label{fig:intro}
\end{figure}

Our goal is to determine the assignment matrix $P$, state aggregation function $u$, and lifting function $v$ to establish the low-dimensional skeleton $G_s$ of the graph $G$. Subsequently, by modeling the dynamics $\frac{dx_{s,i}}{dt}$ on the skeleton and mapping it back to the sub-nodes, we aim to achieve an accurate prediction of the long-term evolution of node states $X_{horizon}$, given history observations $X_{lookback}$. The whole process is illustrated in Figure \ref{fig:intro}.


\section{Method}
\begin{figure*}[!h]
\centering
\includegraphics[width=\textwidth]{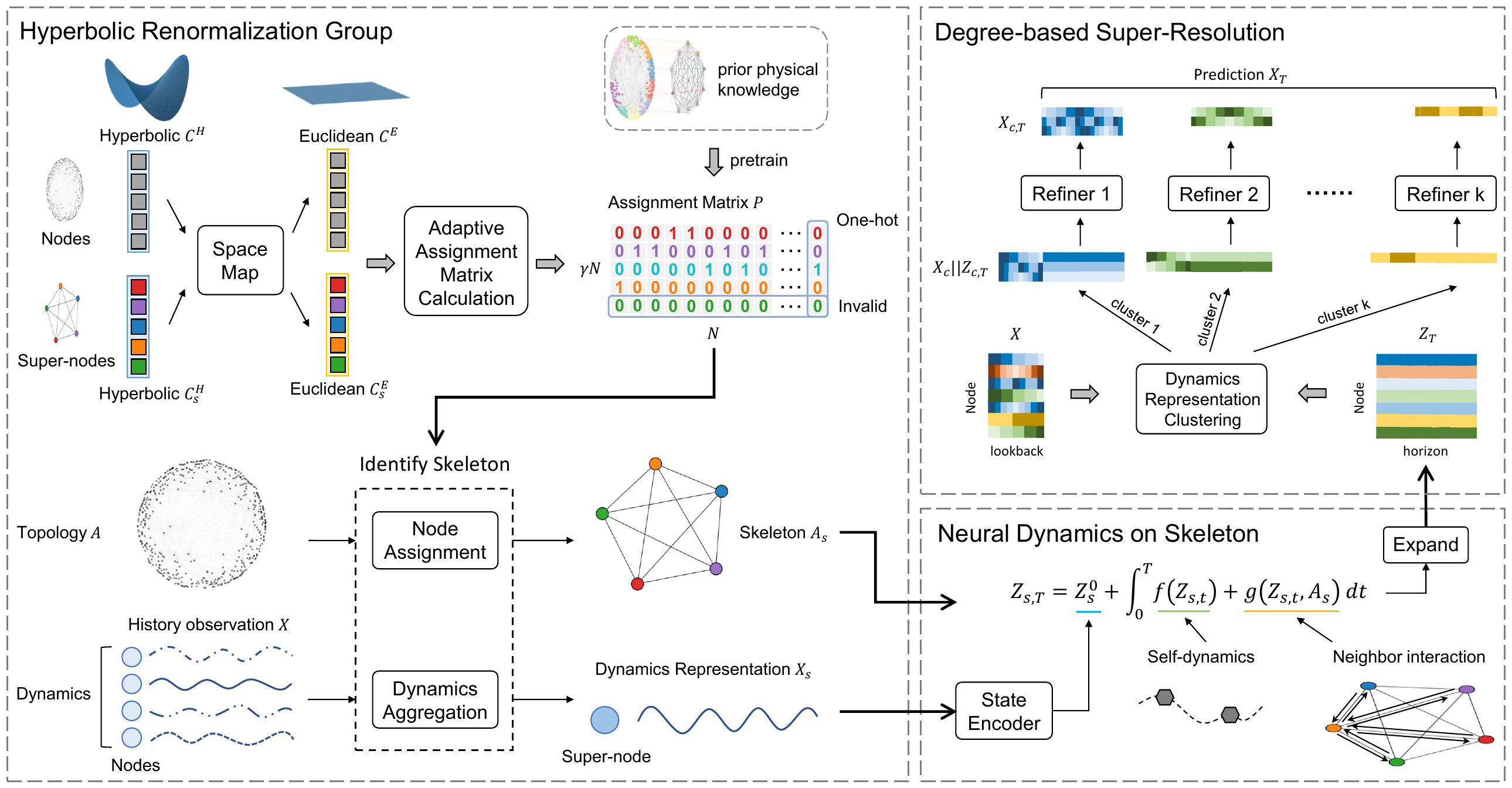}
\caption{Overall framework of DiskNet: (1) Hyperbolic Renormalization Group, which identifies the representation and skeleton of network dynamics; (2) Neural Dynamics on Skeleton, which models the dynamics of super-nodes on the skeleton; and (3) Degree-based Super-Resolution, which lifts the predicted values of super-nodes to the original nodes.}
\label{fig:model}
\end{figure*}

In this section, we propose Dynamics-Invariant Skeleton Neural Network (DiskNet) to predict the long-term dynamics of complex networks via
identifying skeleton in hyperbolic space. The overall framework is illustrated in Figure \ref{fig:model}. According to the challenges mentioned above, we first propose a hyperbolic renormalization group (RG) module which leverages the powerful representation capability of hyperbolic geometry to capture the topological and dynamic similarities of nodes and adaptive calculates the assignment matrix, guiding the node assignment and dynamics aggregation. Then, a neural ODE model is designed to model the neural dynamics of super-nodes on the skeleton, incorporating two mechanisms: self-dynamics and neighbor interactions. Finally, a degree-based clustering super-resolution module is proposed to lift the dynamic representation on the skeleton to the original nodes.

\subsection{Hyperbolic Renormalization Group}
Given the adjacency matrix $A$ of a complex network, the key to identifying the skeleton $A_s$ of the network lies in determining the assignment matrix $P$ based on the topological properties of the network and the latent similarity of node dynamics. We work in the hyperbolic space \cite{liu2019hyperbolic} to measure the similarity between the original nodes and the super-nodes. Based on this, we derive an adaptive assignment matrix to aggregate the dynamic representations of the sub-nodes for each super-node, which joint incorporates features of the topology and dynamics. We introduce prior physical knowledge by initializing the hyperbolic embeddings of the super-nodes to guide model training.

\subsubsection{Learnable hyperbolic embeddings.}
The Poincaré disk is a 2-dimensional hyperbolic geometry where the negative curvature property aligns the hyperbolic distances between node embeddings with their proximity in the graph. Hyperbolic spaces of the same dimension have a larger capacity than Euclidean spaces, which makes it more suitable for representing both the topological and dynamic information of nodes in large-scale networks. The distance between two points on the Poincaré disk is given by 
\begin{equation}
    d^c_H(x,y)=\frac{1}{\sqrt{|c|}}\mathrm{arcosh}(1-\frac{2c||x-y||^2}{(1+c||x||^2)(1+c||y||^2)}) .
\end{equation}

Previous works have focused on learning hyperbolic embeddings $C^H \in \mathbb{R}^{N \times 2}$ for network topology, which is frozen in DiskNet after initialization and used for hierarchical representation of the original topology. Our core idea is to maintain a learnable hyperbolic embedding $C^H_s \in \mathbb{R}^{\gamma N \times 2}$ for each super-node to capture the dynamic characteristics of all its sub-nodes. We first initialize $C^H_s$ according to the method described in Sec. \ref{sec:init} to introduce prior physical knowledge, and then fine-tune it during end-to-end training. Considering that the computational rules defined in Euclidean space do not apply to vectors in hyperbolic space, we project $C^H$ and $C^H_s$ onto corresponding Euclidean coordinates using the logarithmic map
\begin{equation}
    \theta^E_z=\frac{2}{\sqrt{|c|}\lambda^c_z}\mathrm{arctanh}(\sqrt{|c|} ||-z \oplus_c \theta^H||)\frac{-z \oplus_c \theta^H}{||-z \oplus_c \theta^H||} ,
\end{equation}
where $\oplus_c$ means Möbius addition \cite{mathieu2019continuous} and $\lambda^c_x=\frac{2}{1+c||x||^2}$. $\theta^H$ and $\theta^E$ denote the vector in hyperbolic and Euclidean space, respectively. Consistent with most of the work \cite{liu2019hyperbolic, nickel2017poincare, peng2020mix, wang2019hyperbolic}, the tangent space chosen for the hyperbolic manifold is at the origin point $\mathbb{O}$, and the curvature is set as $-1$. We provide detailed verification of the superiority of hyperbolic embeddings for representing both the topological and dynamic similarities of nodes in Sec. \ref{sec:skeleton}.

\subsubsection{Adaptive assignment matrix.}
Based on the hyperbolic embeddings of the original nodes and super-nodes, we compute the adaptive assignment matrix as 
\begin{equation}
    P = {\rm softmax}(\tilde{C}_s\tilde{C}^T),
\end{equation}
where softmax is calculated row-wise, $\tilde{C} = MLP(C^E)$ and $\tilde{C}_s = MLP(C^E_s)$. We minimize $L_E=\frac{1}{N}\sum^N_{i=1}{H(P_i)}$, where $H$ denotes the entropy function and $P_i$ is the $i$-th column of $P$, to constrain each column of the assignment matrix to be close to a one-hot vector. Additionally, we minimize $L_R=||A,P^TP||_F$ to heuristically guide the assignment matrix in preserving the skeleton of the original topology, where $||\cdot||_F$ denotes the Frobenius norm.

\subsubsection{Node dynamics aggregation.}
We employ graph convolutional neural networks to capture the propagation of dynamics in the original nodes and aggregate them into dynamic representations of the super-nodes as 
\begin{equation}
\begin{aligned}
    H &= \sigma(\tilde{D}^{-\frac{1}{2}} \tilde{A} \tilde{D}^{-\frac{1}{2}} X) \Theta_1) \in \mathbb{R}^{N \times d}, \\
    X_s &= PH \in \mathbb{R}^{\gamma N \times d} ,
\end{aligned}
\end{equation}
where $\tilde{D} = \sum_{j}{\tilde{A}_{ij}}$, $\tilde{A} = A + I$ and $\Theta$ is learnable parameters. The connections between super-nodes are described by the adjacency matrix $A_s = PAP^T$, where an element $a_{s,ij}$ indicates the total number of connections between sub-nodes within a super-node ($i=j$) or between different super-nodes ($i \neq j$).

\subsubsection{Physics-informed initialization.} \label{sec:init}
Statistical physics assigns the meaning in terms of node similarity to angular coordinates of the node embeddings in the Poincaré disk and employs the renormalization group based on angular coordinates \cite{garcia2018multiscale}. This design helps preserve the degree distribution and clustering coefficient of the network topology in the skeleton. While this method does not guarantee an effective representation of network dynamics, DiskNet can automatically adjust the learnable hyperbolic embeddings of super-nodes during subsequent end-to-end training. Therefore, we sort the original nodes according to angular coordinates and divide them into groups of $\frac{1}{\gamma}$ nodes to predefine the assignment matrix $P_0$, which is used for pretraining the adaptive assignment module with the loss function of $L_p = \frac{1}{\gamma N^2}\sum{|P-P_0|}$.
We provide the pseudo-code of the pre-training phase in the appendix \ref{apx:pre-train}.
The initial values of hyperbolic embeddings for super-nodes are set as $c^H_{s,i} = \frac{1}{|P_{0,i}|}\sum_{j \in P_{0,i}}{c^H_j}$, where $P_{0,i}$ denotes the set of sub-nodes belonging to the super-node $i$. We extensively analyze the relationship between angular coordinates and node dynamics in Sec. \ref{sec:skeleton}.

\subsection{Neural Dynamics on Skeleton}
After obtaining the skeleton of network dynamics and the aggregated state of super-nodes, we maintain the form of network dynamics as Eqn. \ref{equ:dynamics} and model it using the neural ODE \cite{chen2018neural}.

\subsubsection{ODE function}
We now define the forward ODE function for the latent dynamics $\frac{dZ_s}{dt}$ of the skeleton. Taking into account that the evolution of each super-node $x_{s,i}$ is influenced by its self-dynamics and coupling dynamics, the parameterized time derivative consists of two terms: one donating the self-dynamics function $f(Z_s)$ and another representing the interaction with neighbors $g(Z_s, A_s)$. Thus the dynamic equation for each super-node is given by 
\begin{equation}
    \frac{dZ_s}{dt} = f(Z_s) + g(Z_s, A_s), 
\end{equation}
where $f$ is an MLP and $g$ is a GNN responsible for information propagation
\begin{equation}
    g(Z_s, A_s) = \sigma(A\sigma(Z_s\Theta_3)\Theta_2) .
\end{equation}

\subsubsection{Solve}
Given the ODE function, the predicted trajectory of the skeleton dynamics can be solved by any ODE solver as an initial value problem
\begin{equation}
    Z_{s,T} = Z_{s,0} + \int^T_0{f(Z_{s,t}) + g(Z_{s,t}, A_s) dt} ,
\end{equation}
where the initial state $Z_{s,0} = MLP(X_s) \in \mathbb{R}^{\gamma N \times 1}$. This allows us to predict the state of super-nodes at any continuous time point $T$.

\subsection{Lift to Original Nodes}
After obtaining the predicted trajectories of super-nodes on the skeleton, it is necessary to lift them back to each individual node in the original graph to complete the prediction of network dynamics. However, there is inevitably information loss during the aggregation of super-nodes, such as the heterogeneity of sub-node states within the same super-node. Simply copying the predicted values of super-nodes to their sub-nodes is not sufficient for accurate expanding. To overcome this challenge, inspired by research in statistical physics \cite{ma2023generalized,gao2016universal}, we designed a super-resolution module based on degree clustering.

\subsubsection{Degree-based clustering.}
By employing a degree-weighted mean-field approach, one can compress the state and tipping point of degree-similar nodes into a super-node for prediction \cite{liu2022network,gao2016universal,ma2023generalized}. This implies that the collective dynamics of nodes exhibit homogeneity with respect to their degrees. Inspired by this, considering that real-world networks often follow power-law distributions, we choose to use the logarithm of degrees as features for clustering nodes by K-means algorithm and perform super-resolution within each cluster. In fact, in the Poincaré disk, the degree of a node directly corresponds to its radial coordinate. Therefore, nodes with similar radial coordinates should be identified as belonging to the same super-node. 

\subsubsection{Super-resolution.}
Based on the effective representation of the skeleton dynamics, the design of the super-resolution module does not need to be complex. We denote the set of nodes in the $c$-th cluster as $N_c$, with their historical observations as $X_c$ and predicted values of the corresponding super-node as $Z_{c,T}$. Since the expand operation simply copies the predicted values of the super-node to its sub-nodes without differentiation, $Z_{c,T}$ is a coarse and homogeneous prediction for each node. We refine $Z_{c,T}$ based on the historical observations $X_c$, resulting in accurate prediction as
\begin{equation}
    X_{c,T} = \sigma((h_0 || h_1)\Theta_4) ,
\end{equation}
where $||$ donates the concatenation operation, $h_0 = MLP(X_c)$ and $h_1 = MLP(Z_{c,T})$. Thus, we have completed the prediction of network dynamics. We demonstrate in Sec. \ref{sec:k} that the prediction accuracy rapidly reaches the upper limit as the cluster count $k$ increases. Therefore, a limited setting of $k$ is sufficient to achieve good prediction performance.

\subsection{Training}
We now describe the training process of DiskNet. Before the start of end-to-end training, we first initialize the hyperbolic embeddings of both the original nodes and super-nodes, as described in Sec. \ref{sec:init}, and pretrain the adaptive assignment module leveraging the prior physical knowledge. Subsequently, in each iteration, the model learns the adaptive assignment matrix guided by the loss functions $L_E$ and $L_R$. The prediction of the skeleton dynamics is updated using the L2 error as $L_s = MSE(\overline{Z}_{s,T}, Z_{s,T})$, where $\overline{Z}_{s,T}$ represents the predicted states of the super-node, and $Z_{s,T}$ is the ground truth obtained based on the assignment matrix and the state encoder. Finally, the L2 error of the prediction for all original nodes, $L_p = MSE(\overline{X}_T, X_T)$, is utilized for updating the super-resolution module. Due to the Poincaré disk being a conformal map of Euclidean space, when updating node embeddings with gradients, it is necessary to scale the gradients according to the Riemannian metric tensor
\begin{equation}
    \nabla_H(\theta) = (\frac{1-||\theta||}{2})^2 \nabla_E(\theta) ,
\end{equation}
where $\nabla_E(\theta)$ and $\nabla_H(\theta)$ is the Euclidean and hyperbolic gradient of vector $\theta$, respectively. After completing end-to-end training, we separately fine-tune the super-resolution module to enhance the prediction performance.
\section{Experiments}
In this section, we present the evaluation results of DiskNet. We first describe the dynamics and network topology used for testing, followed by an introduction to the baseline for comparison. Then, we analyze the long-term prediction performance, interpretability of the identified skeleton, and sensitivity analysis. Detailed information about the software and hardware environment for all experiments is provided in the appendix \ref{apx:envir}.

\subsection{Experiment Setup}
We conduct experiments on three representative nonlinear dynamic systems, operating on five real-world network topologies and two synthetic topologies. The details of dataset generation can be found in the appendix \ref{apx:simulation}.

\subsubsection{Network Topology}
We consider five real-world network topologies \cite{rossi2015network} from infrastructure, social networks, brain networks, and the World Wide Web \footnote{https://networkrepository.com/index.php}: 
(a) power grids of the United States (\textit{PowerGrid});
(b) brain networks of \textit{Drosophila};
(c) mutually liked Facebook pages (\textit{Social}); 
(d) pages linking to www.epa.gov (\textit{Web}); 
(e) ties between two non-US-based airports (\textit{Airport}). 
Additionally, we also consider two synthetic networks: 
(f) Barabási–Albert network (\textit{BA}) \cite{barabasi1999emergence}; 
and (g) Watts–Strogatz network (\textit{WS}) \cite{watts1998collective}. 
The statistical characteristics of all topologies are shown in Table \ref{tab:main_result}.

\subsubsection{Network Dynamics}

We considered three representative dynamics from biology and physics: (a) Hindmarsh-Rose dynamics \cite{borges2018inference}, (b) FitzHugh-Nagumo dynamics \cite{fitzhugh1961impulses}, 
 and (c) Coupled Rössler dynamics \cite{arenas2008synchronization}. The governing equations and details of parameter settings are provided in the appendix \ref{apx:dynamics}. 

\begin{table*}[ht]
\renewcommand{\arraystretch}{1.25}
\caption{Average performance of the long-term prediction with 120 time steps with standard deviation
from 10 runs. The best results are highlighted in bold, and the suboptimal results are emphasized with an underline.}
\centering
\resizebox{\textwidth}{!}{%
\begin{tabular}{ll|c|c|c|c|c|c|c}
\hline \hline
 & \multicolumn{1}{l}{} & \multicolumn{1}{c}{PowerGrid} & \multicolumn{1}{c}{Drosophila} & \multicolumn{1}{c}{Social} & \multicolumn{1}{c}{Web} & \multicolumn{1}{c}{Airport} & \multicolumn{1}{c}{BA} & \multicolumn{1}{c}{WS} \\
 & \multicolumn{1}{l}{Nodes} & \multicolumn{1}{c}{5,300} & \multicolumn{1}{c}{1,770} & \multicolumn{1}{c}{3,892} & \multicolumn{1}{c}{4,252} & \multicolumn{1}{c}{2,904} & \multicolumn{1}{c}{5,000} & \multicolumn{1}{c}{5,000} \\
 & \multicolumn{1}{l}{Edges} & \multicolumn{1}{c}{8,271} & \multicolumn{1}{c}{8,905} & \multicolumn{1}{c}{17,239} & \multicolumn{1}{c}{8,896} & \multicolumn{1}{c}{15,644} & \multicolumn{1}{c}{14,991} & \multicolumn{1}{c}{10,000} \\
 & \multicolumn{1}{l}{Avg degree} & \multicolumn{1}{c}{3.121} & \multicolumn{1}{c}{10.062} & \multicolumn{1}{c}{8.859} & \multicolumn{1}{c}{4.184} & \multicolumn{1}{c}{10.774} & \multicolumn{1}{c}{5.996} & \multicolumn{1}{c}{4.000} \\
 & \multicolumn{1}{l}{Avg clustering coeff} & \multicolumn{1}{c}{0.088} & \multicolumn{1}{c}{0.265} & \multicolumn{1}{c}{0.374} & \multicolumn{1}{c}{0.071} & \multicolumn{1}{c}{0.456} & \multicolumn{1}{c}{0.010} & \multicolumn{1}{c}{0.368} \\
  \midrule
 & \multicolumn{1}{l}{MAE $\downarrow$} & \multicolumn{1}{c}{$Mean \ \pm\ Std.$} & \multicolumn{1}{c}{$Mean \ \pm{\ Std.}$} & \multicolumn{1}{c}{$Mean \ \pm{\ Std.}$} & \multicolumn{1}{c}{$Mean \ \pm{\ Std.}$} & \multicolumn{1}{c}{$Mean \ \pm{\ Std.}$} & \multicolumn{1}{c}{$Mean \ \pm{\ Std.}$} & \multicolumn{1}{c}{$Mean \ \pm{\ Std.}$} \\
 \midrule
\multirow{8}{*}{\rotatebox{90}{Hindmarsh-Rose}}
 & DCRNN & $0.3652\ \pm\ 0.0315$ & $0.5616\ \pm\ 0.0392$ & $0.4869\ \pm\ 0.0478$ & $0.4226\ \pm\ 0.0453$ & $0.5457\ \pm\ 0.0434$ & $0.4284\ \pm\ 0.0552$ & $0.3436\ \pm\ 0.0330$ \\
 & GraphWaveNet & $0.6083\ \pm\ 0.0080$ & $0.6300\ \pm\ 0.0011$ & $0.6144\ \pm\ 0.0012$ & $0.6124\ \pm\ 0.0039$ & $0.6448\ \pm\ 0.0064$ & $0.6148\ \pm\ 0.0057$ & $0.6081\ \pm\ 0.0027$ \\
 & AGCRN & $0.2417\ \pm\ 0.0093$ & $0.3590\ \pm\ 0.0078$ & $0.3236\ \pm\ 0.0053$ & $0.2801\ \pm\ 0.0091$ & $0.3591\ \pm\ 0.0064$ & $0.2964\ \pm\ 0.0072$ & $0.2422\ \pm\ 0.0114$ \\
 & NCDN & $0.1554\ \pm\ 0.0027$ & $0.3255\ \pm\ 0.0031$ & $0.2938\ \pm\ 0.0101$ & $0.2298\ \pm\ 0.0092$ & $0.3349\ \pm\ 0.0030$ & $0.2672\ \pm\ 0.0082$ & $0.2002\ \pm\ 0.0065$ \\
 & STGNCDE & $0.3283\ \pm\ 0.0173$ & $0.4836\ \pm\ 0.0161$ & $0.4171\ \pm\ 0.0111$ & $0.3817\ \pm\ 0.0227$ & $0.4802\ \pm\ 0.0081$ & $0.3952\ \pm\ 0.0157$ & $0.3379\ \pm\ 0.0118$ \\
 & MTGODE  & $\textbf{0.1283}\ \pm\ \textbf{0.0045}$ & \underline{$0.3170\ \pm\ 0.0141$} & \underline{$0.2264\ \pm\ 0.0057$} & \underline{$0.1791\ \pm\ 0.0083$} & \underline{$0.3131\ \pm\ 0.0085$} & \underline{$0.1895\ \pm\ 0.0075$} & $\textbf{0.1295}\ \pm\ \textbf{0.0057}$ \\
 & FourierGNN & $0.1651\ \pm\ 0.0053$ & $0.3793\ \pm\ 0.0119$ & $0.3024\ \pm\ 0.0083$ & $0.2378\ \pm\ 0.0112$ & $0.3830\ \pm\ 0.0110$ & $0.2251\ \pm\ 0.0198$ & $0.1601\ \pm\ 0.0073$ \\ 
 & DiskNet & \underline{$0.1290\ \pm\ 0.0029$} & $\textbf{0.2515}\ \pm\ \textbf{0.0124}$ & \textbf{$\textbf{0.2086}\ \pm\ \textbf{0.0052}$} & $\textbf{0.1645}\ \pm\ \textbf{0.0018}$ & $\textbf{0.2769}\ \pm\ \textbf{0.0028}$ & $\textbf{0.1878}\ \pm\ \textbf{0.0027}$ & \underline{$0.1359\ \pm\ 0.0031$} \\
 \midrule
\multirow{8}{*}{\rotatebox{90}{FitzHugh-Nagumo}} 
 & DCRNN & $0.4487\ \pm\ 0.0368$ & $0.5136\ \pm\ 0.0859$ & $0.4780\ \pm\ 0.0993$ & $0.4810\ \pm\ 0.0599$ & $0.5297\ \pm\ 0.0659$ & $0.4809\ \pm\ 0.1139$ & $0.3597\ \pm\ 0.0170$ \\
 & GraphWaveNet & $0.5805\ \pm\ 0.0060$ & $0.5036\ \pm\ 0.0013$ & $0.5523\ \pm\ 0.0006$ & $0.4538\ \pm\ 0.0016$ & $0.5127\ \pm\ 0.0011$ & $0.5223\ \pm\ 0.0026$ & $0.5904\ \pm\ 0.0007$ \\
 & AGCRN & $0.2903\ \pm\ 0.0264$ & $0.2100\ \pm\ 0.0175$ & $0.2272\ \pm\ 0.0105$ & $0.1924\ \pm\ 0.0070$ & $0.2136\ \pm\ 0.0142$ & $0.2130\ \pm\ 0.0122$ & $0.2179\ \pm\ 0.0132$ \\
 & NCDN & $0.4281\ \pm\ 0.0048$ & $0.2750\ \pm\ 0.0058$ & $0.3550\ \pm\ 0.0056$ & $0.2467\ \pm\ 0.0092$ & $0.2965\ \pm\ 0.0046$ & $0.3193\ \pm\ 0.0074$ & $0.4291\ \pm\ 0.0101$ \\
 & STGNCDE & $0.3165\ \pm\ 0.0229$ & $0.2645\ \pm\ 0.0215$ & $0.2534\ \pm\ 0.0168$ & $0.2918\ \pm\ 0.0336$ & $0.2633\ \pm\ 0.0243$ & $0.2532\ \pm\ 0.0127$ & $0.2573\ \pm\ 0.0048$ \\
 & MTGODE  & $0.1420\ \pm\ 0.0133$ & $0.1331\ \pm\ 0.0138$ & $\textbf{0.0826}\ \pm\ \textbf{0.0054}$ & \underline{$0.0873\ \pm\ 0.0034$} & \underline{$0.0934\ \pm\ 0.0052$} & \underline{$0.0906\ \pm\ 0.0039$} & $0.1103\ \pm\ 0.0037$ \\
 & FourierGNN & \underline{$0.1359\ \pm\ 0.0106$} & \underline{$0.1301\ \pm\ 0.0100$} & $0.1315\ \pm\ 0.0116$ & $0.1016\ \pm\ 0.0054$ & $0.1392\ \pm\ 0.0076$ & $0.1171\ \pm\ 0.0168$ & \underline{$0.0677\ \pm\ 0.0091$} \\
 & DiskNet & $ \textbf{0.0997}\ \pm\ \textbf{0.0076}$ & $\textbf{0.0851}\ \pm\ \textbf{0.0021}$ & \underline{$0.0932\ \pm\ 0.0061$} & $\textbf{0.0575}\ \pm\ \textbf{0.0024}$ & $\textbf{0.0917}\ \pm\ \textbf{0.0029}$ & $\textbf{0.0800}\ \pm\ \textbf{0.0037}$ & $\textbf{0.0591}\ \pm\ \textbf{0.0072}$ \\
 \midrule
\multirow{8}{*}{\rotatebox{90}{Coupled Rössler}}
 & DCRNN & $0.5825\ \pm\ 0.0277$ & $0.4927\ \pm\ 0.0235$ & $0.4719\ \pm\ 0.0340$ & $0.4062\ \pm\ 0.0139$ & $0.3977\ \pm\ 0.0415$ & $0.5046\ \pm\ 0.0417$ & $0.4655\ \pm\ 0.0173$ \\
 & GraphWaveNet & $0.4394\ \pm\ 0.0496$ & $0.3743\ \pm\ 0.0329$ & $0.3665\ \pm\ 0.0093$ & $0.2969\ \pm\ 0.0528$ & $0.3465\ \pm\ 0.0088$ & $0.3951\ \pm\ 0.0136$ & $0.3555\ \pm\ 0.0113$ \\
 & AGCRN & $0.1837\ \pm\ 0.0146$ & $0.1562\ \pm\ 0.0065$ & $0.1563\ \pm\ 0.0059$ & $0.1188\ \pm\ 0.0030$ & $0.1603\ \pm\ 0.0065$ & $0.1699\ \pm\ 0.0060$ & $0.1387\ \pm\ 0.0073$ \\
 & NCDN & $0.5908\ \pm\ 0.0145$ & $0.5077\ \pm\ 0.0235$ & $0.4992\ \pm\ 0.0219$ & $0.3743\ \pm\ 0.0130$ & $0.4567\ \pm\ 0.0133$ & $0.4593\ \pm\ 0.0297$ & $0.5104\ \pm\ 0.0177$ \\
 & STGNCDE & $0.2621\ \pm\ 0.0595$ & $0.2021\ \pm\ 0.0362$ & $0.2487\ \pm\ 0.0626$ & $0.1877\ \pm\ 0.0337$ & $0.3004\ \pm\ 0.0641$ & $0.2891\ \pm\ 0.0535$ & $0.1727\ \pm\ 0.0316$ \\
 & MTGODE & \underline{$0.0707\ \pm\ 0.0075$} & $0.0896\ \pm\ 0.0125$ & \underline{$0.0907\ \pm\ 0.0076$} & $0.0753\ \pm\ 0.0125$ & $\textbf{0.0824}\ \pm\ \textbf{0.0104}$ & \underline{$0.0748\ \pm\ 0.0085$} & \underline{$0.0860\ \pm\ 0.0044$} \\
 & FourierGNN & $0.1013\ \pm\ 0.0067$ & \underline{$0.0849\ \pm\ 0.0060$} & $0.1368\ \pm\ 0.0063$ & \underline{$0.0689\ \pm\ 0.0021$} & $0.1070\ \pm\ 0.0040$ & $0.0905\ \pm\ 0.0094$ & $0.1281\ \pm\ 0.0078$ \\
 & DiskNet & $\textbf{0.0612}\ \pm\ \textbf{0.0042}$ & $\textbf{0.0640}\ \pm\ \textbf{0.0032}$ & $\textbf{0.0878} \ \pm\  \textbf{0.0148}$ & $\textbf{0.0680}\ \pm\ \textbf{0.0035}$ & \underline{$0.0909\ \pm\ 0.0111$} & $\textbf{0.0616}\ \pm\ \textbf{0.0116}$ & $\textbf{0.0761}\ \pm\ \textbf{0.0089}$ \\

 \hline \hline
\end{tabular}%
}
\label{tab:main_result}
\end{table*}

\subsection{Baseline}
For all the datasets, we compare with the following GNN-based state-of-the-art methods.
\begin{itemize}
    \item \textbf{DCRNN} \cite{li2017diffusion} employs bidirectional random walks and an encoder-decoder architecture with scheduled sampling to enable accurate spatio-temporal forecasting.
    \item \textbf{GraphWaveNet} \cite{wu2019graph} incorporates an adaptive dependency matrix that captures hidden spatial dependencies through node embeddings. 
    \item \textbf{AGCRN} \cite{bai2020adaptive}: introduces two adaptive modules along with recurrent networks to automatically capture fine-grained spatial and temporal correlations in traffic series.
    \item \textbf{NDCN} \cite{zang2020neural} captures continuous-time dynamics on complex networks by combining neural ODEs and GNNs.
    \item \textbf{STGNCDE} \cite{choi2022graph} combines spatio-temporal graph neural networks with neural controlled differential equations for predicting multi-variate time series.
    \item \textbf{MTGODE} \cite{jin2022multivariate} abstracts multivariate time series into dynamic graphs with time-evolving node features and models their continuous dynamics in the latent space.
    \item \textbf{FourierGNN} \cite{yi2023fouriergnn} is the state-of-the-art GNN model for multi-variate time series forecasting. It introduces hypervariate graphs and performs matrix multiplications in Fourier space.
\end{itemize}

\subsection{Performance Evaluation}
In this section, we analyze the long-term prediction performance, identified skeletons, and computational cost of DiskNet.

\subsubsection{Long-term Prediction}
We evaluated the performance of DiskNet in long-term predictions for all topologies and dynamics based on Mean Absolute Error (MAE). The model was provided with observed trajectories from the past 12 time steps and tasked with predicting the future 120 steps of node states. In all experiments, the reduction ratio $\gamma$ and cluster count $k$ of DiskNet is set to 50\% and 10, respectively, and the reasons for this setting are explained in Sec. \ref{sec:sensitivity}. 

The average performance for 120 time steps are shown in Table \ref{tab:main_result}. It is noteworthy that DiskNet consistently outperforms all baselines by a large margin in long-term prediction performance across almost all scenarios, with the improvement reaching over 50\% in some cases. Furthermore, it can be observed that the suboptimal model varies across different scenarios, even among state-of-the-art models, indicating fluctuations in their performance. This, on the one hand, demonstrates that our approach is capable of accurately identifying and modeling the skeleton of network dynamics in a wide range of scenarios. On the other hand, it emphasizes the significance of the low-dimensional skeleton for stable and accurate long-term predictions. In contrast, the NCDN and MTGODE model, which also employ graph neural ODE, perform well only in a few scenarios, possibly because it models the entire network without effectively aggregating the crucial collective behaviors of nodes.

\begin{figure}[!t]
\centering
\includegraphics[width=0.49\textwidth]{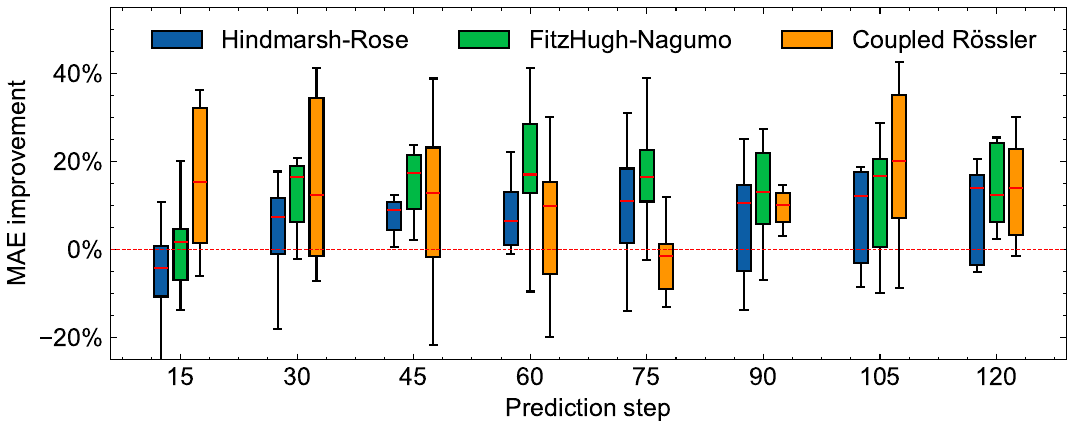}
\caption{Percentage improvement in MAE as a function of the horizon for different dynamics.}
\label{fig:mae}
\end{figure}

We present the detailed statistics of the percentage improvement of DiskNet compared to the suboptimal baseline across all topologies for different dynamics in Figure \ref{fig:mae}, which demonstrates the performance of DiskNet in terms of prediction error at different time steps. In terms of short-term prediction performance, DiskNet is slightly weaker than the suboptimal baseline due to inevitable information loss during the aggregation of node states. However, this short-term information gradually diminishes as the time scale increases, and DiskNet excels in capturing long-term evolution through the skeleton.

\subsubsection{Dynamics Skeleton} \label{sec:skeleton}
We then study the skeleton of the dynamic identified by DiskNet through comparative experiments and visual analysis. We compare five baselines including heuristics, graph pooling and renormalization group: (a) random assignment; (b) threshold assignment based on node degree; (c) threshold assignment based on betweenness centrality; (d) DiffPool \cite{ying2018hierarchical}; (e) GMPool \cite{ko2023grouping}; and (f) static RG \cite{garcia2018multiscale}. The first three methods select super-nodes according to certain rules, and the sub-nodes are assigned randomly to super-nodes. The assignment rules of two graph pooling methods are consistent with the original paper. The RG method assigns nodes solely based on their topological features, thus it is referred to as static RG. In all experiments, to ensure fairness, after obtaining the assignment matrix using the baseline, modules such as state aggregation and skeleton dynamics, are trained end-to-end using DiskNet. The reduction ratio $\gamma$ was set to 50\% for all models.

\begin{figure*}[!t]
\centering
\subfigure[RG (no dynamics)]{\label{fig:static RG}
\includegraphics[width=0.24\textwidth]{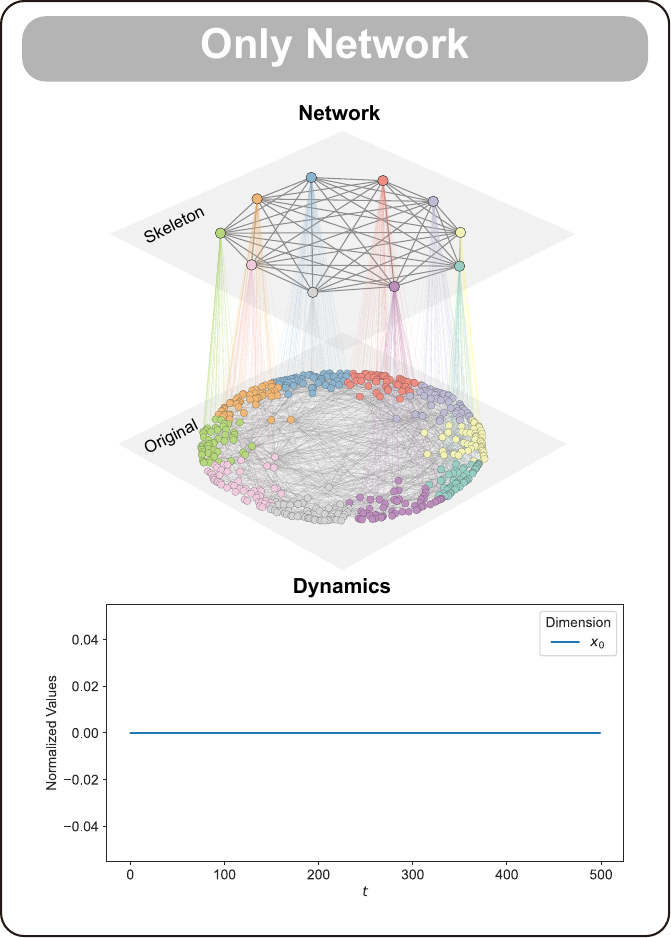}}
\subfigure[DiskNet (Hindmarsh-Rose)]{\label{fig:skeleton_HR}
\includegraphics[width=0.24\textwidth]{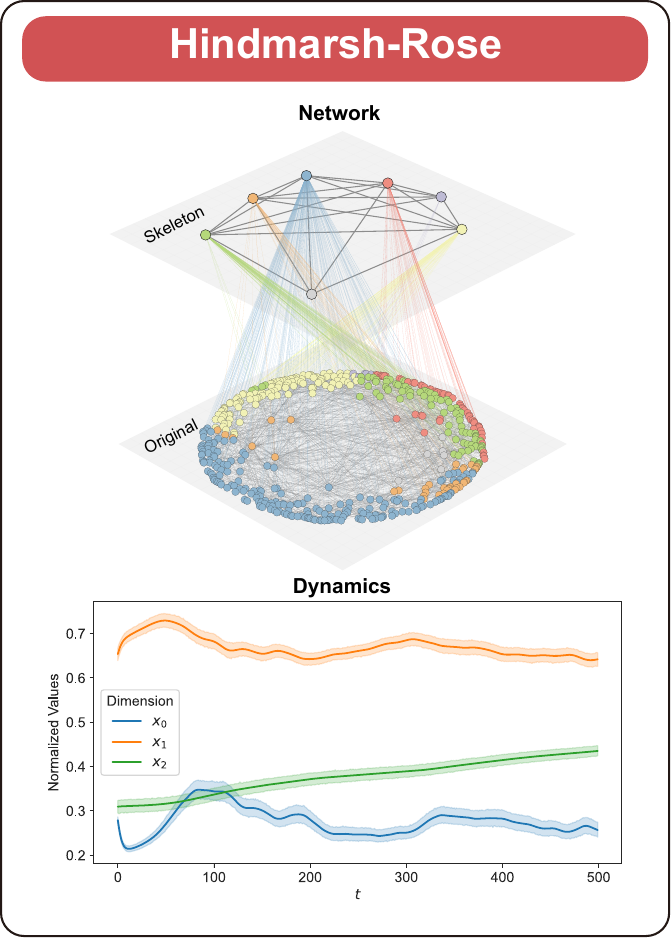}}
\subfigure[DiskNet (FitzHugh-Nagumo)]{\label{fig:skeleton_FHN}
\includegraphics[width=0.24\textwidth]{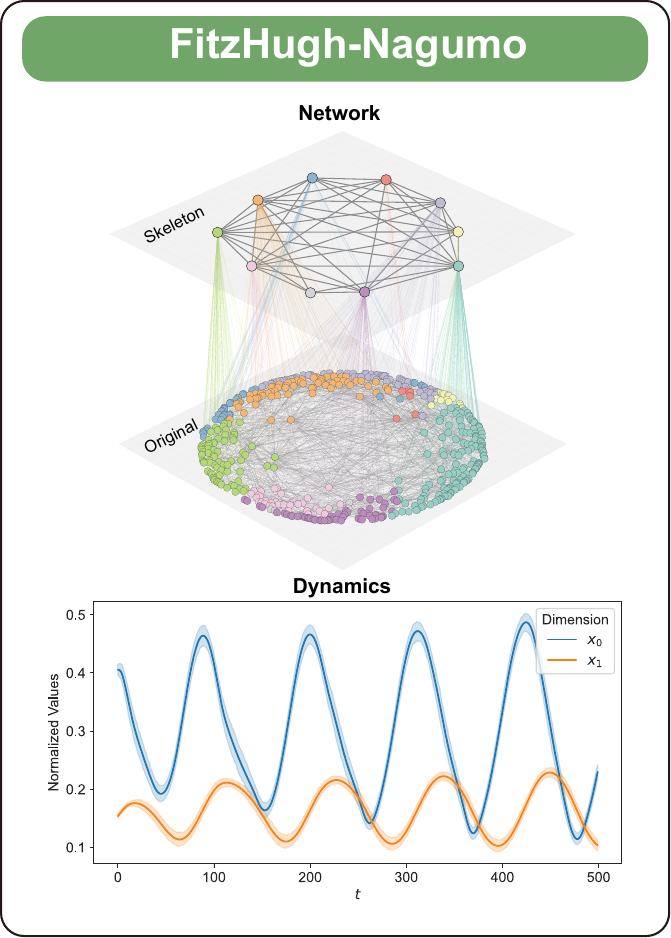}}
\subfigure[DiskNet (Coupled Rössler)]{\label{fig:skeleton_CR}
\includegraphics[width=0.24\textwidth]{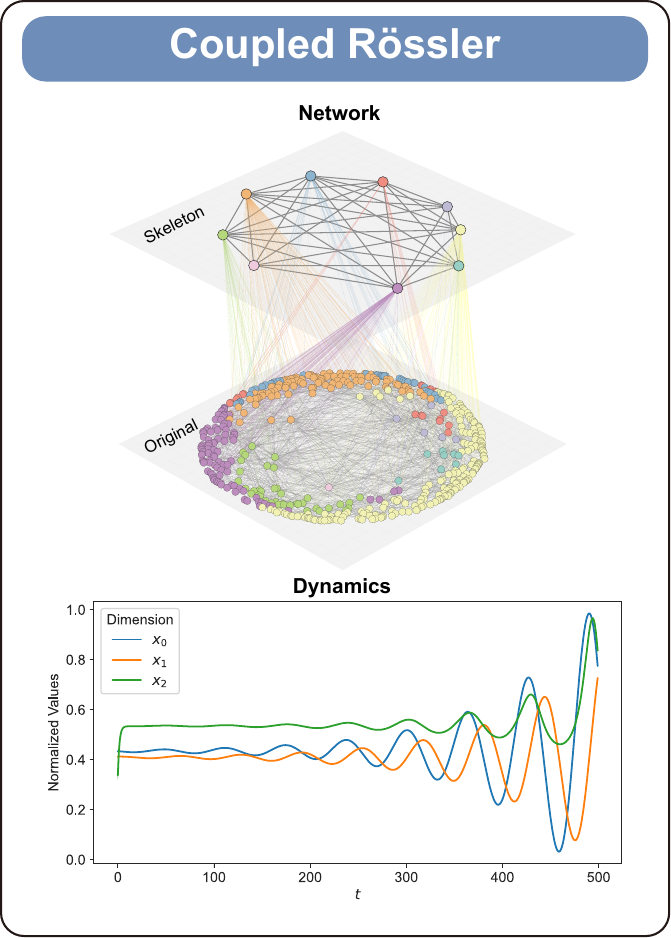}}
\caption{Identified skeleton for different dynamics on the BA-500 networks of DiskNet and static RG model. The bottom half of each figure depicts the average evolution trajectory of all nodes.}
\label{fig:hyper}
\end{figure*}

\begin{table}[!t]
\caption{Comparison of the prediction MAE with 120 time steps for Hindmarsh-Rose dynamics among different skeleton identification methods.}
\centering
\resizebox{\columnwidth}{!}{%
\begin{tabular}{lccccccc}
\midrule
\multicolumn{1}{l}{} & \multicolumn{1}{c}{PowerGrid} & \multicolumn{1}{c}{Drosophila} & \multicolumn{1}{c}{Social} & \multicolumn{1}{c}{Web} & \multicolumn{1}{c}{Airport} & \multicolumn{1}{c}{BA} & \multicolumn{1}{c}{WS} \\
\midrule
Random & 0.1813 & 0.2891 & 0.2497 & 0.1986 & 0.2842 & 0.2169 & 0.1748 \\
Degree & 0.1817 & 0.2895 & 0.2437 & 0.1997 & 0.2902 & 0.2187 & 0.1874 \\
Betweenness & 0.1802 & 0.2948 & 0.2485 & 0.1997 & 0.2897 & 0.2197 & \underline{0.1567} \\
DiffPool & 0.1757 & 0.2863 & 0.2376 & 0.1804 & \underline{0.2785} & 0.2147 & 0.1815 \\
GMPool & 0.1698 & 0.2821 & 0.2299 & \underline{0.1756} & 0.2873 & 0.2069 & 0.1618 \\
static RG & \underline{0.1517} & \underline{0.2640} & \underline{0.2179} & 0.1763 & 0.2796 & \underline{0.1909} & 0.1572 \\
DiskNet & \textbf{0.1290} & \textbf{0.2515} & \textbf{0.2086} & \textbf{0.1645} & \textbf{0.2769} & \textbf{0.1878} & \textbf{0.1359} \\
\midrule
\end{tabular}%
}
\label{tab:skeleton}
\end{table}

We conduct experiments on Hindmarsh-Rose dynamics across all topologies as shown in Table \ref{tab:skeleton}. With the powerful representation capability of hyperbolic space, the skeleton identified by DiskNet captures the long-term dynamics of the nodes, resulting in more accurate predictions compared to the baseline in all scenarios. Heuristic methods, by solely considering the topological characteristics while disregarding the node dynamics, exhibit inferior performance compared to random assignment in certain scenarios. Furthermore, graph pooling methods, which represent node dynamics in Euclidean space to derive the assignment matrix, are unable to effectively measure node similarity and capture latent dynamic interactions as effectively as hyperbolic space. Similar observations have also been seen in node classification tasks \cite{nickel2017poincare,nickel2018learning}. Although the static RG method also utilizes hyperbolic geometry to represent node features, it only considers static topological features for calculating the assignment matrix, disregarding the similarity in node dynamics. As a result, it is unable to identify the skeleton of long-term dynamics.

We visualize the skeletons identified by the static RG method and DiskNet for the same topology but different dynamics in Figure \ref{fig:hyper}, where the dynamic curves represent the average trajectories of all nodes. We utilize the result of a 500-node BA network with reduction ratio $\gamma=0.02\%$ to avoid visual clutter caused by large network size (visualized skeletons for real-world networks in Table \ref{tab:main_result} are provided in the appendix. Since the static RG method does not consider dynamics, the identified skeletons are the same for the three dynamics, as shown in Figure \ref{fig:static RG}. In contrast, DiskNet shows variations in the identified skeletons for different dynamics. For FitzHugh-Nagumo dynamics (Figure \ref{fig:skeleton_FHN}) and Coupled Rössler dynamics (Figure \ref{fig:skeleton_CR}), where all nodes exhibit oscillatory behavior, DiskNet identifies a similar number of super-nodes, and the assignment patterns are also similar. However, for the Hindmarsh-Rose dynamics, which have heterogeneous node dynamics but a simple trend, the corresponding skeleton tends to be captured with fewer super-nodes.

Further analysis of the skeletons identified by DiskNet reveals that the majority of super-nodes share the same local neighborhood with their sub-nodes, but there are also instances where the positions of sub-nodes deviate, especially in the case of Hindmarsh-Rose dynamics (Figure \ref{fig:skeleton_HR}). While the existing work \cite{garcia2018multiscale} has demonstrated that nodes with similar angles on the Poincaré disk should belong to the same super-node when considering only topological characteristics, DiskNet suggests that this conclusion no longer holds when considering node dynamics. Even nodes with substaintially different angular coordinates may be assigned to the same super-node due to similar dynamic properties. Furthermore, it is observed that nodes with the same angular coordinates may be assigned to different super-nodes due to differences in radial coordinates, which is attributed to the correlation between node degrees and their dynamics \cite{ma2023generalized}. Finally, it is worth noting that some super-nodes are only assigned a small number of sub-nodes, which is further explained in Sec. \ref{sec:gamma}.

\subsubsection{Computational Cost} \label{sec:time}

\begin{figure}[!t]
\centering
\includegraphics[width=0.49\textwidth]{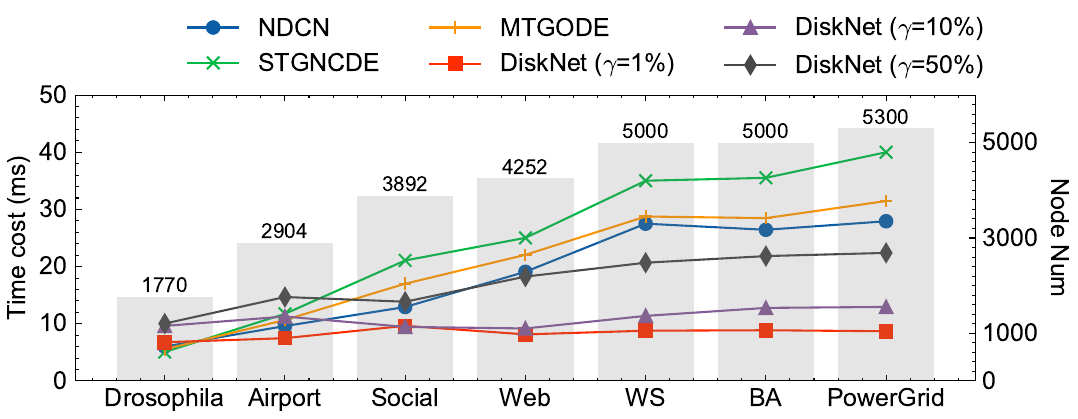}
\caption{Time cost per iteration on different network topologies for Hindmarsh-Rose dynamics.}
\label{fig:time}
\end{figure}

DiskNet only requires integration of the forward ODE function on the skeleton during inference, which significantly saves computational resources compared to previous neural ODE-based methods. We compare DiskNet with varying reduction ratio $\gamma$ and two other neural ODE-based baselines on different topologies to measure the time cost for single-sample inference. The results in Figure \ref{fig:time} demonstrate that even with the additional computational overhead from the super-resolution module, DiskNet with $\gamma=50\%$ achieves performance comparable to NDCN. In fact, as the network size increases, the dimension reduction potential of network dynamics often exceeds 50\% (which we validate in Sec. \ref{sec:gamma}). Therefore, DiskNet has higher practical value than previous neural ODE-based methods in predicting large-scale network dynamics.

\subsubsection{Physics-informed Initialization} \label{sec:init}

The performance of DiskNet in Table \ref{tab:main_result} are based on the physics-informed initialization, which allows the model to start learning the optimal parameters from a good initial point. We evaluate versions of DiskNet on the Coupled Rössler dynamics that (1) do not initialize the hyperbolic embeddings of super-nodes and/or (2) do not pre-train the assignment matrix calculation module. The results (in Table \ref{tab:init}) show that while DiskNet without physical initialization performs competitively compared to the baseline in many scenarios, it is always inferior to the physically initialized version, confirming the necessity of physics-informed initialization.

\begin{table}[!t]
\caption{Comparison of the prediction MAE with 120 time steps for Coupled Rössler dynamics among different versions of DiskNet. We do not initialize the hyperbolic embedding and pre-training, denoted as $w/o\ Emb$ and $w/o\ PT$.}
\centering
\resizebox{\columnwidth}{!}{
\begin{tabular}{lccccccc}
\midrule
\multicolumn{1}{l}{} & \multicolumn{1}{c}{PowerGrid} & \multicolumn{1}{c}{Drosophila} & \multicolumn{1}{c}{Social} & \multicolumn{1}{c}{Web} & \multicolumn{1}{c}{Airport} & \multicolumn{1}{c}{BA} & \multicolumn{1}{c}{WS} \\
\midrule
DiskNet & \textbf{0.0612} & \textbf{0.0640} & \textbf{0.0878} & \textbf{0.0680} & \textbf{0.0909} & \textbf{0.0616} & \textbf{0.0761} \\
w/o Emb & \underline{0.1166} & \underline{0.1014} & 0.1269 & \underline{0.0864} & 0.1172 & 0.1118 & 0.1262 \\
w/o PT & 0.1297 & 0.1018 & 0.1299 & 0.0879 & \underline{0.1160} & \underline{0.0911} & \underline{0.0814} \\
w/o Emb \& PT & 0.1377 & 0.1080 & \underline{0.1020} & 0.0910 & 0.1215 & 0.1149 & 0.0838 \\
\midrule
\end{tabular}
}
\label{tab:init}
\end{table}

\subsection{Sensitivity Analysis} \label{sec:sensitivity}
In this section, we analyze the impact of two important hyperparameters, namely the reduction ratio $\gamma$ and the clustering count $k$, on the prediction performance of DiskNet.

\subsubsection{Reduction ratio $\gamma$} \label{sec:gamma}
\begin{figure}[!t]
\centering
\subfigure[BA]{
\includegraphics[width=0.22\textwidth]{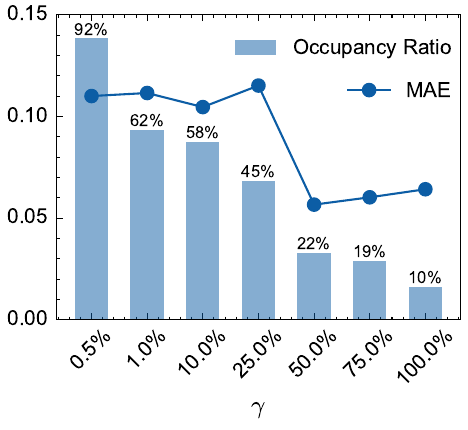}}
\hspace{0.01\textwidth}
\subfigure[PowerGrid]{
\includegraphics[width=0.22\textwidth]{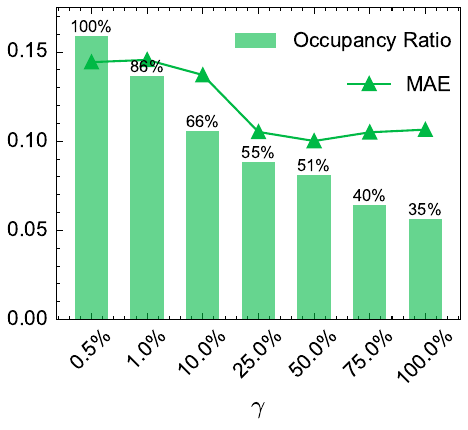}}
\caption{MAE and Occupancy ratio as functions of $\gamma$ for Coupled Rössler dynamics on PowerGrid and BA networks.}
\label{fig:gamma}
\end{figure}

The reduction ratio $\gamma$ sets an upper limit on the number of super-nodes in the skeleton. Although the model automatically determines the assignment relationship based on the topological and dynamics properties, allowing some super-nodes unassigned to any sub-nodes, the proportion of super-nodes does not exceed $\gamma$. We conducted tests on the Coupled Rössler dynamics in both the BA network and PowerGrid network, as shown in Figure \ref{fig:gamma}. We define the occupancy ratio as the proportion of super-nodes that are assigned at least one sub-node out of the total available super-nodes. It is observed that when $\gamma$ is less than 1\%, the prediction performance is the worst, and the occupancy ratio is the highest. This indicates that the upper limit of super-nodes is too low to identify an appropriate skeleton. After $\gamma$ exceeds 25\%, the prediction performance becomes similar. This suggests both the low-dimensional characteristics of network dynamics and indicates that setting $\gamma$ to 50\% is generally adequate, as the model is capable of adaptively determining the appropriate occupancy ratio. Finally, although both networks consist of approximately 5,000 nodes, the occupancy ratio of the PowerGrid network is slightly higher than that of the BA network for the same $\gamma$. This suggests that the Coupled Rössler dynamics in the PowerGrid network has relatively higher dimensions and pose greater prediction challenges, which corresponds to the prediction errors in Table \ref{tab:main_result}.

\subsubsection{Cluster count $k$} \label{sec:k}
\begin{figure}[!t]
\centering
\subfigure[Social MSE]{\label{fig:k1}
\includegraphics[width=0.22\textwidth]{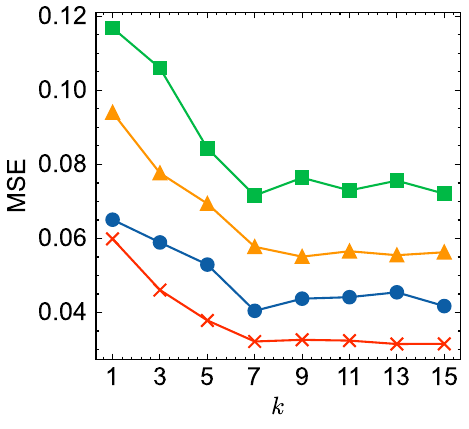}}
\hspace{0.01\textwidth}
\subfigure[Web MSE]{\label{fig:k2}
\includegraphics[width=0.22\textwidth]{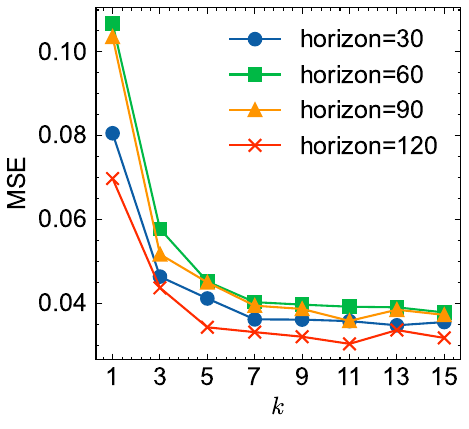}}
\caption{MSE of different horizons as a functions of $k$ for FitzHugh-Nagumo dynamics on Social and Web networks.}
\label{fig:k}
\end{figure}

The cluster count $k$ represents the number of refiners in the super-resolution module and is positively correlated with the prediction performance. However, it needs to be balanced with the parameter size of the model. We conduct prediction tests at different time scales (horizon) on the FitzHugh-Nagumo dynamics in both the Social network and Web network, which is shown in Figure \ref{fig:k}. When $k$ is less than 7, increasing $k$ leads to lower prediction errors. However, beyond $k$ of 7, further increasing $k$ does not significantly improve the performance. Therefore, we recommend using the default value of 10 for $k$ in general, which balances both performance and computational cost.

\section{Related Work}

\subsection{Learning Network Dynamics}

The dynamics of many real-world systems, characterized by nonlinearity and multi-scale properties \cite{li2023learning}, are often abstracted into complex network models, thereby illustrating the interactions between nodes \cite{barzel2015constructing,gao2016universal,liu2024empowering}. With the development of deep learning techniques such as GNNs, data-driven modeling of complex network dynamics has garnered significant attention \cite{ding2024artificial}. Murphy et al. \cite{murphy2021deep} proposed a GNN architecture that accurately models disease propagation on networks with minimal assumptions about the dynamics. NCDN \cite{zang2020neural} combined neural ODE \cite{chen2018neural} and GNNs to model continuous-time dynamics of complex networks for the first time. Huang et al. \cite{huang2021coupled,huang2023generalizing} successfully modeled dynamic topology and cross-environment network dynamics by introducing the edge dynamics ODE and environment encoders, respectively. However, despite the demonstration \cite{prasse2022predicting, thibeault2024low} that network dynamics evolve in a remarkably low-dimensional subspace, existing approaches still model the evolution of dynamics over the entire topology, which fails to capture the low-dimensional structures that dominate the long-term evolution of the network.

\subsection{Dimension Reduction of Complex Network}

Driven by the high costs of storage and computational expenses in large-scale networks, Serrano et al. \cite{serrano2009extracting} proposed the disparity filter to ensure the retention of small-scale components during the pruning process of multiscale networks. García-Pérez et al. \cite{garcia2018multiscale} designed the hyperbolic renormalization group method based on the assumption of angular similarity to identify the low-dimensional skeleton of complex networks and Villegas et al. \cite{villegas2023laplacian} proposed a Laplacian renormalization group difusion-based picture for heterogeneous networks. Kumar et al. \cite{kumar2023featured} used node feature consistency as guidance and learned the low-dimensional skeleton through end-to-end training. Additionally, some works on motif learning \cite{islam2023mpool,yan2023learning} treat each motif in the original graph as a super-node, thus constructing a skeleton that aggregates higher-order geometric characteristics.
However, these methods only consider the static properties of networks and cannot guide the modeling of network dynamics. Gao et al. \cite{gao2016universal} captured macroscopic dynamics by mapping the original system to an effectively representative one-dimensional system and further developed a degree-based dimensionality reduction method \cite{ma2023generalized} that successfully captures the critical point of phase transitions in heterogeneous networks. However, these traditional expert knowledge-based methods cannot automatically identify the most suitable skeleton based on the collective behavior exhibited of nodes for different topologies and dynamics.

\subsection{Hyperbolic Graph Learning}

The negative curvature property of hyperbolic geometry leads to nonlinear distance metrics, enabling it to better capture the hierarchical relationships and nonlinear dependencies in graph-structured data. Nickel et al. \cite{nickel2017poincare,nickel2018learning} embedded symbolic data into the Poincaré ball and Lorentz model, capturing the hierarchical structure and similarity of unstructured data.  Liu et al. \cite{liu2019hyperbolic} and  Chami et al. \cite{chami2019hyperbolic} introduced the hyperbolic geometry into graph representation learning through Riemannian manifold mapping, significantly improving the performance of node classification tasks. Yang et al. \cite{yang2021discrete, yang2022hyperbolic} further utilized hyperbolic geometry to model the hierarchical structure of time-varying networks. Although these works have demonstrated the suitability of hyperbolic geometry for representing network structure data, there is still a lack of its application in modeling complex network dynamics.

\section{Conclusions and Future Work}

In this paper, we explore the low-dimensional skeleton of complex network dynamics and its crucial role in long-term prediction through a skeleton identification approach based on hyperbolic node embeddings. 
By representing the collective dynamics behavior of nodes in the hyperbolic space, we compute the adaptive assignment matrix to identify the skeleton. Then, we achieve effective long-term prediction by modeling the dynamics of super-nodes and lifting them to the original nodes. Extensive experimental results demonstrate that our model outperforms the baseline with a significant margin in terms of accuracy and robustness. 
In addition, we validate the superior capability of hyperbolic geometry over Euclidean space in representing the low-dimensional structure of network dynamics. We demonstrate the limitations of solely identifying skeletons based on topological characteristics through visualization results, which showcase distinct skeletons resulting from different dynamics on the same topology. 
The analysis results indicate that, while ensuring prediction accuracy, there is rich potential for dimensionality reduction in the dynamics of different types of large-scale networks. In the future, we plan to explore how the topology and dynamics of complex networks affect the maximum achievable dimensionality reduction while ensuring prediction accuracy. Additionally, we aim to design a mechanism for automatically selecting the reduction ratio $\gamma$.




\bibliographystyle{ACM-Reference-Format}
\balance

\bibliography{main}

\balance
\appendix




\section{Datasets}

\subsection{Network Dynamics} \label{apx:dynamics}

\textbf{Hindmarsh-Rose dynamics} \cite{borges2018inference} simulates the spiking activity of neurons in the brain during information processing, with the following governing equations
\begin{equation}\label{HR}
\left\{
    \begin{aligned}
        \frac{d{x_{i,1}}}{dt} &= x_{i,2}-ax^{3}_{i,1}+bx^2_{i,1}-x_{i,3}+I_{ext}+\epsilon(V_{syn}-x_{i,1})\sum^{N}_{j=1}{A_{ij}\mu(x_{j,1})} ,\\
        \frac{d{x_{i,2}}}{dt} &= c-ux^2_{i,1}-x_{i,2} ,\\
        \frac{d{x_{i,3}}}{dt} &= r[s(x_{i,1}-x_0)-x_{i,3}] ,
    \end{aligned}
\right.
\end{equation}
where parameters $a=1$, $b=3$, $c=1$, $u=5$, $s=4$, $r=0.005$, $x_0=-1.6$, coupling strength $\epsilon=0.15$, $V_{syn}=2$, $\lambda=10$, $\Omega_{syn}=1$, and external current is $I_{ext}=3.24$ for all neurons. The coupling term is
\begin{equation}\label{HR_mu}
    \mu(x_{j,1}) = \frac{1}{1+e^{[-\lambda(x_{j,1}-\Omega_{syn})]}} .
\end{equation}

\textbf{FitzHugh-Nagumo dynamics} \cite{fitzhugh1961impulses} is mainly used for describing the activation and deactivation dynamics of spiking neurons. The equations governing the dynamics are
\begin{equation}\label{FHN}
\left\{
    \begin{aligned}
        \frac{d{x_{i,1}}}{dt} &= x_{i,1}-x^3_{i,1}-x_{i,2}-\epsilon\sum^{N}_{j=1}{A_{ij}\frac{x_{j,1}-x_{i,1}}{k^{in}_{i}}} ,\\
        \frac{d{x_{i,2}}}{dt} &= a+bx_{i,1}+cx_{i,2} ,
    \end{aligned}
\right.
\end{equation}
where parameters $\epsilon=1$, $a=0.28$, $b=0.5$, $c=-0.04$, and $k^{in}_{i}$ is the in-degree of neuron $i$.

\textbf{Coupled Rössler dynamics} \cite{arenas2008synchronization} describes the dynamics of $n$ oscillators with natural frequencies following a normal distribution, in which the self-dynamics are heterogeneous
\begin{equation}\label{CR}
\left\{
    \begin{aligned}
        \frac{d{x_{i,1}}}{dt} &= -w_{i}x_{i,2}-x_{i,3}+\epsilon\sum^{N}_{j=1}{A_{ij}(x_{j,1}-x_{i,1})} ,\\
        \frac{d{x_{i,2}}}{dt} &= w_{i}x_{i,1}+ax_{i,2} , \\
        \frac{d{x_{i,3}}}{dt} &= b+x_{i,3}(x_{i,1}+c) ,
    \end{aligned}
\right.
\end{equation}
where parameters $\epsilon=0.15$, $a=0.2$, $b=0.2$, $c=-6$, and frequencies $w \sim \mathcal{N}(1,\sigma)$ with $\sigma=0.1$.

\subsection{Simulation} \label{apx:simulation}
We employ Euler's method to numerically solve the dynamics of the aforementioned systems for each topology, generating evolutionary trajectories as the dataset with a time step of 0.01 seconds. The Hindmarsh-Rose model is simulated for 20 seconds, while the FitzHugh-Nagumo and Coupled Rössler models are simulated for 50 seconds, respectively. All simulation trajectories are downsampled at equal intervals to obtain 500 observations. The dataset is then divided into training, validation, and testing sets in a ratio of 6:2:2.

\section{Software and Hardware Environment} \label{apx:envir}

We implement DiskNet in PyTorch and employ the open-source available implementations with default parameters for baselines. All experiments were conducted on the NVIDIA GeForce RTX 4090 GPU. For all datasets, we set the batch size to 8 and trained for 50 epochs with a learning rate of 0.001.

\section{Pseudo-code of the pre-training phase} \label{apx:pre-train}

We perform supervised pre-training on the Assignment Matrix Calculation module as described in Algorithm \ref{alg::pretrain}. The supervisory signal is given by the assignment matrix $P_0$ using the RG method from statistical physics.

\begin{algorithm}[h]
\caption{Physics-informed Pre-training}
\label{alg::pretrain} 
\begin{algorithmic}[1]
    \Require
        hyperbolic embedding $C^H$ and $C^H_s$;
    \Ensure
        pre-trained MLPs with weights $\theta_1$ and $\theta_2$
    \State calculate assignment matrix $P_0$ by physical knowledge;
    \State convert ($C^H$, $C^H_s$) to ($C^E$, $C^E_s$) using equation (3);
    \For{i=1; i $\le$ n; i++}
        \State node representation $\tilde{C}=MLP_{\theta_1}(C^E)$;
        \State super-node representation $\tilde{C}_s=MLP_{\theta_2}(C^E_s)$;
        \State assignment $P=\mathrm{softmax}(\tilde{C}_s\tilde{C}^T)$;
        \State loss $L_p=\frac{1}{\gamma N}\sum|P-P_0|$;
        \State compute gradient directions $\nabla L_p$;
        \State update $\theta_1$ and $\theta_2$;
    \EndFor
\end{algorithmic}
\end{algorithm}

\end{document}